# Disorder-induced soliton transmission in nonlinear photonic lattices

Yaroslav V. Kartashov, Victor A. Vysloukh, and Lluis Torner

*ICFO-Institut de Ciencies Fotoniques, and Universitat Politecnica de Catalunya, Mediterranean Technology Park, 08860 Castelldefels (Barcelona), Spain*
*Corresponding author: Yaroslav.Kartashov@icfo.es*

We address soliton transmission and reflection in nonlinear photonic lattices embedded into uniform Kerr nonlinear media. We show that by introducing disorder into the guiding lattice channels, one may achieve soliton transmission even under conditions where regular lattices reflect the input beam completely. In contrast, in the parameter range where the lattice is almost transparent for incoming solitons, disorder may induce a significant reflection.

Soliton propagation in quasi-periodic and random refractive index landscapes has been extensively discussed over the last years [1-4]. Under appropriate conditions, disorder may completely change the evolution of light. It may cause the formation of spatially localized states even in linear lattices [5-9]. In the fully nonlinear regime, disorder may result in trapping of strongly localized walking solitons [10] or, vice versa, it may facilitate the transport of such states in quasi-periodic lattice [11]. Disorder leads to Brownian soliton motion [12,13] and percolation [14]. Most of previous studies address the effect of disorder on light evolution in infinite periodic systems, while the effects that may become available in finite disordered systems have not been studied yet.

In this Letter we consider soliton transmission and reflection by disordered nonlinear photonic lattices of finite width embedded into a uniform medium. Transmission in such lattices may be obtained by tailoring solutions in the uniform medium and in the lattice using continuity conditions. Thus, waves in the uniform medium should be matched with Bloch-like lattice eigenfunctions. Since disorder drastically affects the topology of lattice eigenfunctions one expects considerable disorder-induced modifications in the transmission and reflection properties. Backscattering experiments have been already used to study Anderson localization in strongly scattering media, such as semiconductor powders [15,16]. Here we show that addition of disorder into finite lattices may lead to considerable transmission even under conditions where the corresponding regular lattice reflects the input beam completely, or, vice-versa, one may increase reflection under conditions where when the regular lattice is transparent.

We use the nonlinear Schrödinger equation to describe the propagation of light beams inside a disordered optical lattice imprinted in a Kerr nonlinear medium:

$$i\frac{\partial q}{\partial \xi} = -\frac{1}{2}\frac{\partial^2 q}{\partial \eta^2} - pQ(\eta)q - |q|^2 q, \quad (1)$$

Here $\eta$ and $\xi$ are the normalized transverse and longitudinal coordinates, respectively (the second transverse coordinate is omitted because the beam is supposed to be highly elliptical or guided by a slab waveguide); the parameter $p$ characterizes the linear refractive index modulation depth; the lattice shape is described by the function $Q(\eta) = \sum_{m=-(n-1)/2}^{(n-1)/2} G(\eta - \eta_m)$, where the profiles of individual guides are given by $G(\eta) = \exp(-\eta^6 / a^6)$. The coordinates $\eta_m$ of waveguide centers, or scattering centers, are randomized such that $\eta_m = md + s_m$, where $d$ is the spacing of waveguides in the regular lattice, while $s_m$ stands for the random shift of the waveguide center uniformly distributed within the segment $[-S_d, +S_d]$. Such array with a finite number of guides $n$ is embedded into an unbounded uniform Kerr nonlinear medium. We set $p = 11$, $a = 0.3$, and $d = 1.6$ that correspond to a refractive index contrast $\delta n \approx 10^{-3}$ at the wavelength $\lambda = 0.8$ $\mu$m and for a waveguide width of 3 $\mu$m and a mean spacing of 16 $\mu$m.

To gain intuitive insight, we first conducted a qualitative analysis where guides may be considered as scatterers characterized by complex amplitude reflection coefficients $\rho_1$. Within the Born approximation of weak scatterers the net amplitude reflection coefficient for plane waves $\exp(i\alpha\eta)$ interacting with an array of $n$ guides can be estimated as $\rho_n = \rho_1 \sum_{m=1}^{n} \exp(-i\varphi_m)$, where $\varphi_m$ is the phase difference for waves backscattered by the first and $m$-th waveguides. In the regular array where $\varphi_m = 2md\alpha$ the energy reflection coefficient is therefore given by $r_n = |\rho_n|^2 = r_1 \sin^2(n\alpha d) / \sin^2(\alpha d)$, with $r_1 = |\rho_1|^2$. If the Bragg condition $d\alpha = \pi M$ ($M = 1, 2, ...$) is satisfied one gets a considerable reflection with $r_n \sim n^2$. In contrast, in a disordered array the mean value of energy reflection coefficient is given by $\langle r_n \rangle = r_1 n + r_1 \sum_{m,l=1; m \neq l}^{n} \langle \exp[i(\varphi_l - \varphi_m)] \rangle$, where the angular brackets stand for statistical averaging and $\varphi_m, \varphi_l$ are random phases. In particular, in the strong disorder limit, when $\varphi_l - \varphi_m$ fluctuates between $-\pi$ and $\pi$ for different array realizations, the last term vanishes and the mean reflection coefficient $\langle r_n \rangle \simeq r_1 n$ decreases substantially. Such a simple approach works well for plane waves but it is not accurate for localized beams, since only scatterers covered by the beam are involved in the reflection process. The presence of nonlinearity further complicates the picture since it causes beam reshaping upon propagation. Therefore, we integrate Eq. (1) directly to study beam propagation in disordered lattices accurately.

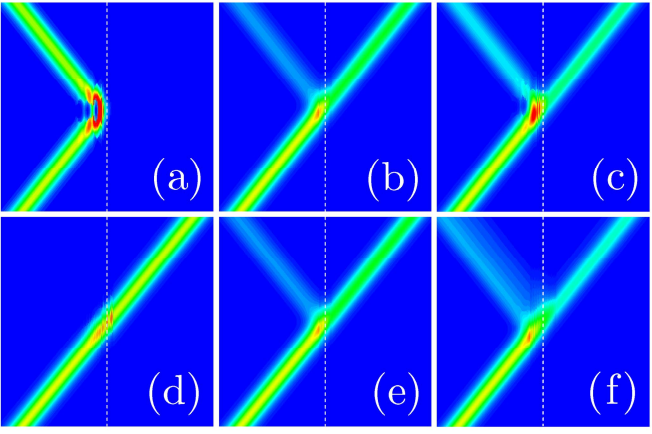

Figure 1. The averaged (over $10^3$ realizations) propagation dynamics of broad solitons with $\chi = 0.05$ colliding with disordered arrays of waveguides at (a) $\alpha = 5.26$, $S_d = 0.00$, (b) $\alpha = 5.26$, $S_d = 0.30$, (c) $\alpha = 5.26$, $S_d = 0.42$, (d) $\alpha = 4.80$, $S_d = 0.00$, (e) $\alpha = 4.80$, $S_d = 0.30$, and (f) $\alpha = 5.26$, $S_d = 0.30$. In (a)-(e) $n = 15$, while in (f) $n = 31$. Dashed lines indicate the mean positions of the central waveguide in the array.

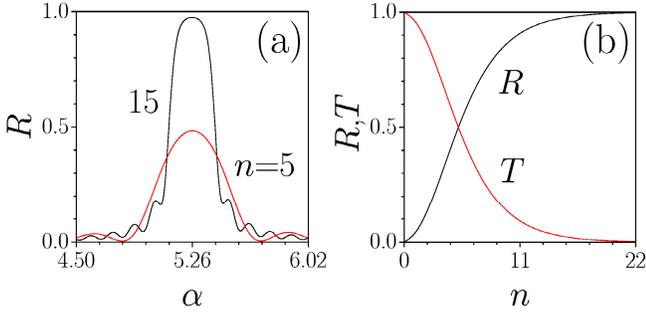

Figure 2. (a) Reflection coefficient in the regular arrays versus input angle for different number of waveguides. (b) Reflection and transmission coefficients versus number of waveguides at $\alpha = 5.26$. In all cases the input soliton has form-factor $\chi = 0.05$.

In the simulations we use as the input the soliton $q|_{\xi=0} = \chi \operatorname{sech}[\chi(\eta - \eta_0)] \exp[i\alpha(\eta - \eta_0)]$ having a form-factor $\chi$ and launched into a uniform nonlinear medium far from the finite array. Here $\alpha$ stands for the incidence angle measured from the array edge in the $(\eta, \xi)$ plane. We calculate statistically averaged reflection $R = \langle U_{\text{ref}} \rangle / U_{\text{in}}$ and transmission $T = \langle U_{\text{tr}} \rangle / U_{\text{in}}$ coefficients, where the reflected energy flow for a particular disorder realization is defined as $U_{\text{ref}} = \int_{-\infty}^{0} |q(\eta, \xi_{\text{end}})|^2 d\eta$, while the transmitted energy flow is defined as $U_{\text{tr}} = \int_{0}^{+\infty} |q(\eta, \xi_{\text{end}})|^2 d\eta$, where the final distance $\xi_{\text{end}} = 2|\eta_0|/\alpha$ was selected to ensure sufficiently large separation of the output reflected and transmitted beams from the array edges. $U_{\text{in}} = 2\chi$ in the input soliton energy flow. The level of disorder is controlled by the parameter $S_d$.

In regular lattices the transmission and reflection coefficients for broad low-power solitons depend strongly on the incidence angle $\alpha$ [17]. While for certain angles total reflection occurs [see propagation dynamics in Fig. 1(a) for $\alpha = 5.26$], even moderate detuning of $\alpha$ from the resonant values results in almost complete soliton transmission [see Fig. 1(d) for $\alpha = 4.80$]. This is illustrated in Fig. 2(a), that shows the angular dependencies of the reflection coefficient for different number of waveguides $n$. Note the qualitative similarity of these dependencies and the expression for the reflection coefficient $\sim \sin^2(n\alpha d)/\sin^2(\alpha d)$ obtained in the Born approximation. The maxima in $R(\alpha)$ are caused by the bandgap lattice spectrum. Thus, reflection occurs when the propagation constant of the input beam, given by $b = -\alpha^2/2$ at $\chi \to 0$ falls into one of the finite gaps of the lattice spectrum. For the parameters of Fig. 2 the $b$ value corresponding to $\alpha = 5.26$ falls exactly into the middle of the third finite gap at $b \in [-14.45, -13.20]$. The function $R(\alpha)$ exhibits several maxima associated with different gaps. The maximum reflection coefficient increases with the number of waveguides in the lattice [Fig. 2(b)].

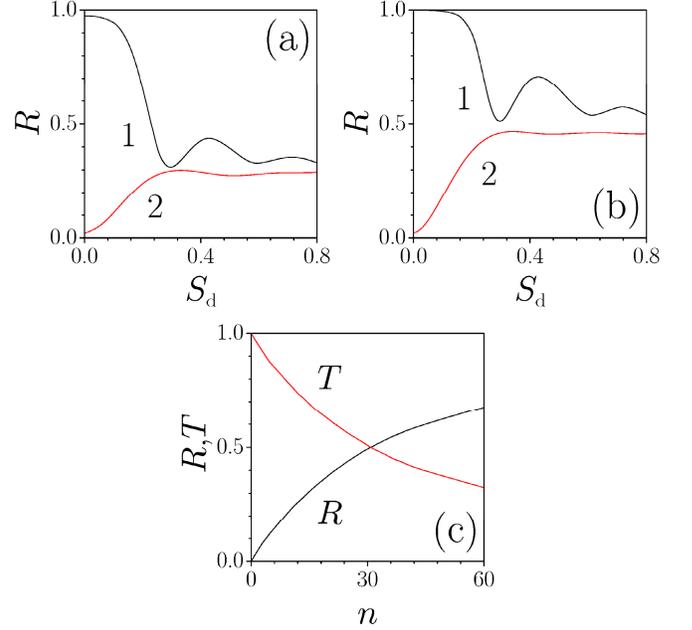

Figure 3. (a) Reflection coefficient versus disorder level for $n = 15$ at $\alpha = 5.26$ (curve 1) and $\alpha = 4.80$ (curve 2). (b) Reflection coefficient versus disorder level for $n = 31$ at $\alpha = 5.26$ (curve 1) and $\alpha = 4.68$ (curve 2). (c) Reflection and transmission coefficients vs $n$ at $\alpha = 5.26$, $S_d = 0.30$. In all cases $\chi = 0.05$.

The important finding reported in this Letter is that the above picture may change drastically in the presence of disorder. Thus, addition of even weak disorder with $S_d = 0.3$ largely increases the mean transmission coefficient $T$ of a lattice that otherwise reflects all light launched at the resonant angle $\alpha = 5.26$ [compare Fig. 1(b) showing the averaged reflection dynamics of disordered lattices with Fig. 1(a) corresponding to regular lattices].

Interestingly, larger disorder does not necessarily result in larger transmission: Fig. 1(c) illustrates that a further growth of disorder up to $S_d = 0.42$ increases the average reflection on the lattice. Under appropriate conditions, disorder may also result in the opposite effect. Namely, it can suppress soliton transmission completely. Thus, Fig. 1(e) shows the averaged dynamics for $S_d = 0.3$ and $\alpha = 4.80$ where one can clearly see the appearance of considerable reflection. This is in complete contrast to Fig. 1(d), where for the identical incident angle solitons propagate through the regular lattice without any appreciable reflection. When the number of waveguides grows, reflection increases too and so does the width of the statistically averaged reflected beam [Fig. 1(f)]. The beam becomes asymmetric and it may have long exponentially decaying tails which are typical of the diffusive regime of light scattering.

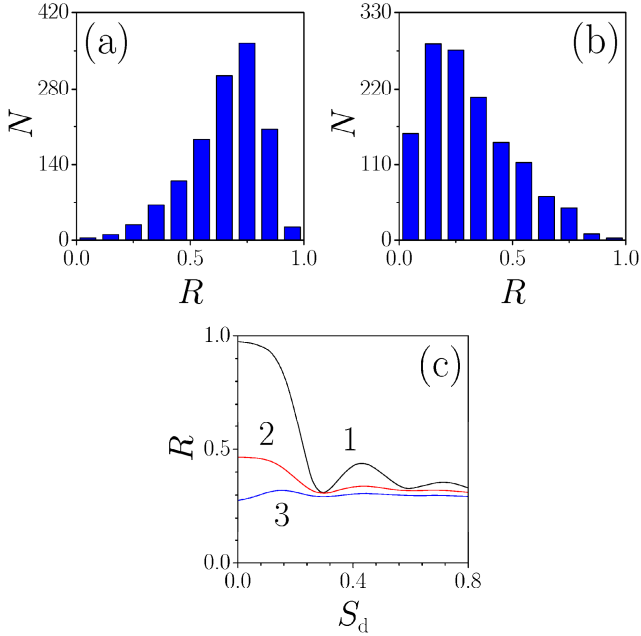

Figure 4. Histograms of the reflection coefficient calculated for 1300 realizations of disordered waveguide arrays for $\chi=0.05$, $n=15$, $\alpha=5.26$ at $S_{\rm d}=0.2$ (a) and $0.3$ (b). (c) Reflection coefficient versus disorder level for $n=15$, $\alpha=5.26$ at $\chi=0.05$ (curve 1), $0.50$ (curve 2), and $1.00$ (curve 3).

The statistical properties of the soliton transmission are analyzed in Figs. 3(a) and 3(b). For incident angles corresponding to resonant reflection (curves 1) the statistically averaged reflection coefficient $R$ first decreases, thus indicating disorder-induced partial transparency, and then starts performing decaying oscillations, gradually approaching a limiting value at large values of $S_{\rm d}$. We did not increase disorder beyond $d/2$ because for $S_{\rm d}>0.8$ the neighboring guides may overlap (note that already at $S_{\rm d}>0.4$ the width of eigenmodes in the disordered lattice becomes smaller than the array width). In contrast, for $\alpha$ values detuned from the resonant angle (curves 2) one observes an almost monotonic growth of the reflection coefficient with increasing $S_{\rm d}$, indicating disorder-induced partial reflection. For all values of the input angle $\alpha$ the reflection coefficient $R$ approaches the same limiting value at strong disorders. Such limiting value grows as the number of waveguides increases. For example, for $n=15$ the limiting value amounts to about $0.3$, while for $n=31$ it approaches $0.5$. This indicates that the largest variations of the reflection coefficient are available in arrays with small $n$, while the largest variations of the transmission are possible in arrays with large $n$. In any case, the first deep minimum of $R(S_{\rm d})$ is attained at the relatively small disorder $S_{\rm d}=0.3$. The dependencies of the reflection and transmission coefficients on $n$ are presented in Fig. 3(c). The plots show a much slower growth of $R$ in comparison to regular lattices [see Fig. 2(b)].

Figures 4(a) and 4(b) show histograms of the reflection coefficient calculated over $1300$ random lattice realizations. Both histograms are strongly asymmetric, indicating a considerable difference between mean and most probable values of $R$. At the small disorder $S_{\rm d}=0.2$ reflection scenarios dominate, but already at $S_{\rm d}=0.3$ transmission becomes more probable.

Nonlinearity substantially affects the reflection properties of the lattice [see Fig. 4(c) that depicts the average reflection coefficient versus $S_{\rm d}$ for different form-factors $\chi$ of input soliton]. The impact of nonlinearity is most pronounced in weakly disordered arrays with $S_{\rm d}<0.3$ where addition of nonlinearity results in drastic reduction of reflection even in the regular case. This occurs due to the reduction of reflection on individual guides due to nonlinearity-mediated diminishing of refractive index contrast and to the overall decrease of the number of scatterers covered by the soliton. Interestingly, at $S_{\rm d}=0.3$ when disorder results in maximal diminishing of reflection coefficient in the linear case, the coefficient $R$ is almost insensitive to variations in $\chi$.

Summarizing, the disorder-induced soliton transmission and reflection reported here illustrate another example of the new phenomena that appear due to the interplay between disorder and nonlinearity. Here we focused on predictions for the averaged quantities, which may be tested in sets of waveguide arrays that may be fabricated (see, e.g., [8]) or created by optical lattice induction, a technique specially suited to address statistical problems [6]. It is worth stressing, however, that our predictions have key implications beyond statistics because the disorder-induced effects may occur for every sample.

References

1) F. Abdullaev, Theory of Solitons in Inhomogeneous Media (Wiley, New York, 1994).
2) F. Abdullaev and J. Garnier, Prog. Opt. **48**, 35 (2005).
3) F. Lederer, G. I. Stegeman, D. N. Christodoulides, G. Assanto, M. Segev, and Y. Silberberg, Phys. Rep. **463**, 1 (2008).
4) Y. V. Kartashov, V. A. Vysloukh and L. Torner, Prog. Opt. **52**, 63 (2009).
5) T. Pertsch, U. Peschel, J. Kobelke, K. Schuster, H. Bartelt, S. Nolte, A. Tünnermann, and F. Lederer, Phys. Rev. Lett. **93**, 053901 (2004).
6) T. Schwartz, G. Bartal, S. Fishman and M. Segev, Nature **466**, 52 (2007).
7) Y. Lahini, A. Avidan, F. Pozzi, M. Sorel, R. Morandotti, D. N. Christodoulides, and Y. Silberberg, Phys. Rev. Lett. **100**, 013906 (2008).
8) A. Szameit, Y. V. Kartashov, P. Zeil, F. Dreisow, M. Heinrich, R. Keil, S. Nolte, A. Tünnermann, V. A. Vysloukh, and L. Torner, Opt. Lett. **35**, 1172 (2010).
9) G. Kopidakis, S. Komineas, S. Flach, and S. Aubry, Phys. Rev. Lett. **100**, 084103 (2008).
10) Y. V. Kartashov and V. A. Vysloukh, Phys. Rev. E **72**, 026606 (2005).
11) A. Sukhorukov, Phys. Rev. Lett. **96**, 113902 (2006).
12) Y. V. Kartashov, V. A. Vysloukh, and L. Torner, Phys. Rev. A **77**, 051802(R) (2008).
13) V. Folli and C. Conti, Phys. Rev. Lett. **104**, 193901 (2010).
14) Y. V. Kartashov, V. A. Vysloukh, and L. Torner, Opt. Express **15**, 12409 (2007).
15) D. S. Wiersma, P. Bartolini, A. Lagendij, and R. Righini, Nature **390**, 671 (1997).
16) F. J. P. Schuurmans, M. Megens, D. Vanmaekelbergh, and A. Lagendijk, Phys. Rev. Lett. **83**, 2183 (1999).

17) Y. V. Kartashov, V. A. Vysloukh, and L. Torner, Opt. Express **14**, 1576 (2006).

# References with titles

1) F. Abdullaev, Theory of Solitons in Inhomogeneous Media (Wiley, New York, 1994).
2) F. Abdullaev and J. Garnier, "Optical solitons in random media," Prog. Opt. **48**, 35 (2005).
3) F. Lederer, G. I. Stegeman, D. N. Christodoulides, G. Assanto, M. Segev, and Y. Silberberg, "Discrete solitons in optics," Phys. Rep. **463**, 1 (2008).
4) Y. V. Kartashov, V. A. Vysloukh and L. Torner, "Soliton shape and mobility control in optical lattices," Prog. Opt. **52**, 63 (2009).
5) T. Pertsch, U. Peschel, J. Kobelke, K. Schuster, H. Bartelt, S. Nolte, A. Tünnermann, and F. Lederer, "Nonlinearity and disorder in fiber arrays," Phys. Rev. Lett. **93**, 053901 (2004).
6) T. Schwartz, G. Bartal, S. Fishman and M. Segev, "Transport and Anderson localization in disordered two-dimensional photonic lattices," Nature **466**, 52 (2007).
7) Y. Lahini, A. Avidan, F. Pozzi, M. Sorel, R. Morandotti, D. N. Christodoulides, and Y. Silberberg, "Anderson localization and nonlinearity in one-dimensional disordered photonic lattices," Phys. Rev. Lett. **100**, 013906 (2008).
8) Szameit, Y. V. Kartashov, P. Zeil, F. Dreisow, M. Heinrich, R. Keil, S. Nolte, A. Tünnermann, V. A. Vysloukh, and L. Torner, "Wave localization at the boundary of disordered photonic lattices," Opt. Lett. **35**, 1172 (2010).
9) G. Kopidakis, S. Komineas, S. Flach, and S. Aubry, "Absence of wave packet diffusion in disordered nonlinear systems," Phys. Rev. Lett. **100**, 084103 (2008).
10) Y. V. Kartashov and V. A. Vysloukh, "Anderson localization of solitons in optical lattices with random frequency modulation," Phys. Rev. E **72**, 026606 (2005).
11) A. Sukhorukov, "Enhanced soliton transport in quasiperiodic lattices with introduced aperiodicity," Phys. Rev. Lett. **96**, 113902 (2006).
12) Y. V. Kartashov, V. A. Vysloukh, and L. Torner, "Brownian soliton motion," Phys. Rev. A **77**, 051802(R) (2008).
13) V. Folli and C. Conti, "Frustrated Brownian motion of nonlocal solitary waves," Phys. Rev. Lett. **104**, 193901 (2010).
14) Y. V. Kartashov, V. A. Vysloukh, and L. Torner, "Soliton percolation in random optical lattices," Opt. Express **15**, 12409 (2007).
15) D. S. Wiersma, P. Bartolini, A. Lagendij, and R. Righini, "Localization of light in a disordered medium," Nature **390**, 671 (1997).
16) F. J. P. Schuurmans, M. Megens, D. Vanmaekelbergh, and A. Lagendijk, "Light scattering near the localization transition in macroporous GaP networks," Phys. Rev. Lett. **83**, 2183 (1999).
17) Y. V. Kartashov, V. A. Vysloukh, and L. Torner, "Bragg-type soliton mirror," Opt. Express **14**, 1576 (2006).